# Experimental measurements of fundamental and high-order spoof surface plasmon polariton modes on ultrathin metal strips

Hong Xiang[1], Qiang Zhang[2], Jiwang Chai[1], Fei Fei Qin[2], Jun Jun Xiao[2], and Dezhuan Han [1]*

[1] *Department of Applied Physics, Chongqing University, Chongqing 400044, China*

[2] *College of Electronic and Information Engineering, Shenzhen Graduate School, Harbin Institute of Technology, Shenzhen 518055, China*

*E-mail: dzhan@cqu.edu.cn

**Abstract**

Propagation of spoof surface plasmon polaritons (spoof SPPs) on comb-shaped ultrathin metal strips made of aluminum foil and printed copper circuit are studied experimentally and numerically. With a near field scanning technique, electric field distributions on these metal strips are measured directly. The dispersion curves of spoof SPPs are thus obtained by means of Fourier transform of the field distributions in the real space for every frequency. Both fundamental and second order modes are investigated and the measured dispersions agree well with numerical ones calculated by the finite element method. Such direct measurements of the near field characteristics provide complete information of these spoof SPPs, enabling full exploitation of their properties associated with the field confinement in a subwavelength scale.

## Introduction

Surface plasmon polariton (SPP) is a kind of surface wave propagating along the metal-dielectric interface with wavelengths smaller than that of the incident wave in free space [1,2], which is a consequence of coupling in between electromagnetic (EM) waves and collective oscillations of free electrons in metal. With the subwavelength nature, the EM field of SPP decays exponentially in the normal direction and exhibits strong confinement. The field confinement in the sub-wavelength scale suggests the possibility to go beyond the diffraction limit [3] and enhanced signals of Raman scattering remarkably [4], leading to extensive applications in photonic devices, near field microscopy, light generation, and bio-photonics etc. [2, 5-9]

At low frequencies far away from the optical band, metal behaves like a perfect electric conductor (PEC) and the surface modes on the metal surface become Zenneck [10] or Sommerfeld [11] waves with poor confinement. However, if sub-wavelength structures such as holes in two dimensional (2D) case or grooves in one dimensional (1D) case are introduced in a PEC, EM waves can penetrate into the metal since the holes or grooves can support EM modes. This structured PEC behaves like an effective medium in the long wavelength region with a dispersion of the dielectric function resembling that of a Drude-type metal in the optical frequencies. Therefore the structured PEC may support surface microwaves or THz waves just as the conventional metal does at the optical frequencies [12-17]. As is well known, the plasma frequency in Drude model depends on the density of free electrons, in contrast, the dispersion of the structured PEC is determined by the structure and therefore can be easily tuned by geometric parameters. This kind of surface state supporting on the structured PEC is named as "designer"

SPPs or "spoof" SPPs.

Spoof SPPs are proposed firstly in bulk metal materials with subwavelength structures in three dimensional (3D) geometry. Recently, a concept of conformal surface plasmons (CSPs) [18-20] that propagate on textured ultra-thin metal strips is introduced. These CSPs are realized by the dimension reduction from 3D to 2D and provide much flexibility for applications since CSPs can propagate along the textured ultra-thin metal strip almost freely even it is bent, twisted or folded.

In the study of spoof SPPs in grooved structures and CSPs, most attentions are paid to the fundamental modes, since they are the closest analogues to the "real" SPPs with very similar (even identically shaped) dispersion relationship. Further researches show that with certain geometry parameters, high-order modes [21-24] of spoof SPPs exist in addition to the fundamental modes, which is intrinsically different from the real SPPs at optical frequencies.

A typical structure supporting CSPs, namely, the ultrathin comb-shaped metal strips have been studied in previous works, such as numerical calculation of dispersion curves [18, 23], experimental measurements of energy transmissions [23] and electric field distributions at certain frequency [18]. However, experimental measurements to investigate the eigen-modes in the ultrathin comb-shaped metal strips, such as the dispersion relations, and field distributions of high order modes, are still lacking. To systematically study the propagation properties of the CSPs, one of the most important issues is to figure out the dispersion relations, not only numerically but also experimentally. These information are very important to understand the behavior of CSPs from the practical point of view.

In this paper, systematic experimental results on spoof SPPs propagating on two different kinds of ultra-thin comb-shaped metal strips, i.e., aluminum foil and copper printed circuit, are presented, including the field distributions and dispersion relations. The field distributions are measured by a near-field scanning technique, and the wave numbers hence the dispersion curves are obtained by the Fast Fourier transform (FFT) of the field distributions. Our experimental results agree well with numerical simulations.

**Theory of Spoof SPPs**

Firstly, we give a short review on the general properties of the spoof SPPs propagating on periodically corrugated metallic plates and their corresponding comb-shaped ultra-thin metal strips (dimension reduction from 3D to 2D). Fig. 1(a) shows the cross section of a periodically corrugated metallic plate schematically. The geometry of this structure is characterized by the period *d*, groove width *a*, and groove depth *h*. This figure can also be regarded as a side view of the comb-shaped ultra-thin metal strip with the same geometry as a sectional plane of the periodically corrugated metallic plate.

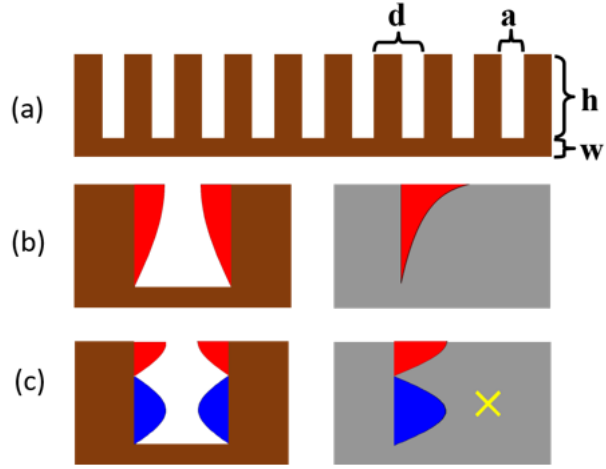

Fig. 1. (a) Cross section and geometry parameters of a corrugated metal plate. (b) Schematic field pattern of the fundamental mode of spoof SPP in the groove (left panel), and field pattern of the "real" SPP inside the metal (right panel). (c) Schematic field pattern of the 2nd order mode of spoof SPP in the groove (left panel), the cross symbol indicates there's no such an analogue of field pattern for the "real" SPPs (right panel).

As the size of the grooves is much smaller than the operating wavelength, the grooved metallic layer can be treated as an effective medium. With a mode expansion method [15, 21], the dispersion relationship of spoof SPP on a periodically corrugated metallic plate reads

$$\beta^2 = \left(\frac{a}{d}\right)^2 k_0^2 \tan^2(k_0 h) + k_0^2 \qquad (1)$$

where $\beta$ is the wavevector or propagating constant, $k_0=\omega/c$ is the corresponding wavevector in free space. At low frequencies, the fundamental modes can be observed and therefore a single dispersion curve can be obtained. If there also exists a high order mode, a cutoff frequency should appear probably above the asymptotic frequency of the fundamental mode. The dispersion for the high order mode starts from the cutoff frequency wherein the corresponding wavevector lies on the light line exactly, namely, $\beta_c=k_{0,c}$. According to Eq. (1), at the cutoff point, $\tan(k_0 h)=0$, which leads to

$$\omega_c = \frac{m\pi c}{h}, \quad \beta_c = \frac{m\pi}{h}. \qquad (2)$$

Here $m$ is non-negative integer. As $\beta$ ranges from $-\pi/d$ to $\pi/d$ in the first Brillouin zone, we have

$$h > md. \qquad (3)$$

Eq. (3) is the condition of the existence of the $(m+1)$-th order mode. Note that the fundamental mode is the first mode with $m=0$ here. From Eq. (3), it is obvious that the fundamental mode corresponding to $m=0$ always exist which has been extensively studied in previous works.

For the comb-shaped ultra-thin metal strips, no analytical calculations have been developed so far. However, numerical simulations based on finite element method (FEM) showed that they have similar dispersion relations except that their dispersion curves have a slight redshift with respect to those of the corresponding periodically corrugated metallic plate.

In Fig. 1(b) and (c), the electric field patterns of spoof SPPs and "real" SPPs are schematically shown. Compared to the SPPs at optical frequencies, the fundamental mode of spoof SPPs behaves similarly to the real SPPs in at least two aspects: (1) the dispersion curves are both below the light line; and (2) electric fields decay to zero monotonically on both sides of the interface. However, for the high-order modes of spoof SPPs, the electric fields on the metal grooves' side do not decay monotonically, instead, the field amplitude can oscillate and have nodes, avoiding decaying to zero directly. For the "real" SPPs, there is no correspondence to such a high order modes of spoof SPPs. This remarkable difference stems from the fact that the corrugated metal plates have metal grooves served as cavities supporting multiple EM resonant modes which is absent in the real SPPs' case.

**Sample fabrication**

To study the fundamental and the 2$^{nd}$ order spoof SPPs modes experimentally, two series of comb-shaped ultra-thin metal strips are fabricated. The geometry parameters are chosen according to Eq. (3) and the operating frequency band is up to 20 GHz. One series of samples are made of aluminum foils with thickness of $t = 0.05\ mm$ with a laser cutter. For all these samples, $d = 5\ mm$, $a = 2\ mm$ and the width of the strip base $w = 3\ mm$. To investigate the different orders of spoof SPPs, four samples with different groove depth $h = 5\ mm$, 9 $mm$, 10 $mm$, 11 $mm$ are fabricated. The other series of samples are fabricated by the standard printed circuit board (PCB) process in which a layer of copper with thickness $t = 0.035\ mm$ is paved on a plastic substrate. The substrate is made of FR-4 with a dielectric constant about 4.2, and its thickness is around 0.6 $mm$. The corresponding structure and geometry are kept exactly the same as those of the aluminum foil samples. Photos of the top view of two samples are shown in the insets in Fig. 2 with the upper one for PCB and the lower one for aluminum foil.

**Experimental setup and measurements**

To measure the EM field distribution point by point, a near-field scanning method is employed. The system comprises a vector network analyzer (VNA, Agilent 5232A), two monopole antennas, and a motor-driven stage. The sample sheet under test is placed on the motor-driven stage with the long side in *x* direction, and short side in *y* direction. One monopole is closely placed at one end of the sample sheet to excite spoof SPP waves. As the near-field generated by the monopole antenna contains a continuous distribution of wavevector components, the condition of phase matching for a surface wave excitation can be satisfied in general. Another monopole antenna is placed on top of the sample with a vertical distance of 1 *mm* from the sample plane to detect the $E_z$ component of the electric field. The VNA is responsible for feeding the source antenna and recording the signals (both amplitude and phase) from the detector antenna. In our experiment, the sample is moved with the stage in both *x* direction and *y* direction with 1 *mm* for each step. In such a way, the field distributions on the whole sample are recorded with a resolution of 1 *mm*.

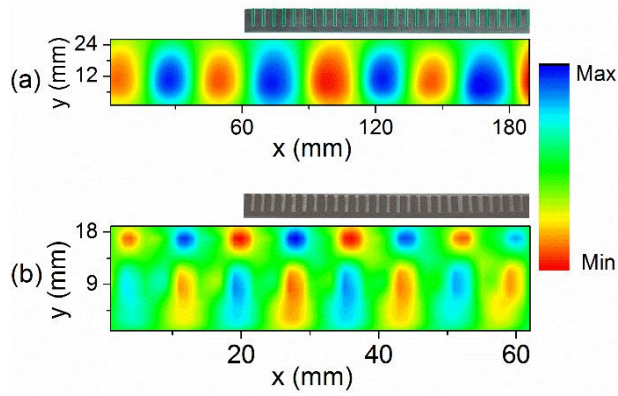

Fig. 2. Electric fields measured on top of the sample with a vertical distance of 1 *mm*. (a) Fundamental mode of the PCB sample with *h*=11 *mm* at 3.5 GHz. (b) 2$^{nd}$ order mode at 14.4 GHz for the aluminum foil sample with *h*=9 *mm*. The insets are photos of two samples. The upper one is for PCB and the lower one is for aluminum foil (free standing). The base and the top of the combs lies in *y*=7 *mm* and 18 *mm* for (a), and *y*=8 *mm* and 17 *mm* for (b).

Fig. 2 shows the measured electric field distributions at two different frequencies as examples. Fig. 2(a) is for the PCB sample with $h = 11\,mm$ at $f = 3.5\,\text{GHz}$, and Fig. 2(b) is for the aluminum foil sample with $h = 9\,mm$ at $f = 14.4\,\text{GHz}$. It can be clearly seen from Fig. 2(a) that the electric amplitude oscillates along the propagating direction (*x*-axis) and decayed to zero along the depth direction of grooves (*y*-axis). While in Fig. 2(b), new character is observed that a node of field amplitude appears along the *y* direction before the amplitude decays to zero. It is easy to identify that Fig. 2(a) belongs to the fundamental mode and Fig. 2(b) belongs to the 2$^{nd}$ order mode.

Based on the measured electric distributions, the dispersion relationship can be obtained by means of Fourier transform. As shown in Fig. 3, the profile of the field along the propagating direction (*x* direction) is extracted and then a fast Fourier transform is conducted. The highest peak in the Fourier spectrum is responsible for the spatial frequency of the spoof SPPs, namely, the wave number $1/\lambda_{spoof\ SPP}$. Other peaks with smaller amplitude are resulted from the subwavelength structure of the samples and scattering of imperfections in the fabrication process. More specially, when the detecting antenna is very close to the sample surface, the details of the surface structures are resolved by the near field probe. Hence at low frequencies, dark lines are found in the field patterns corresponding to the groove area where metal layer is absent.

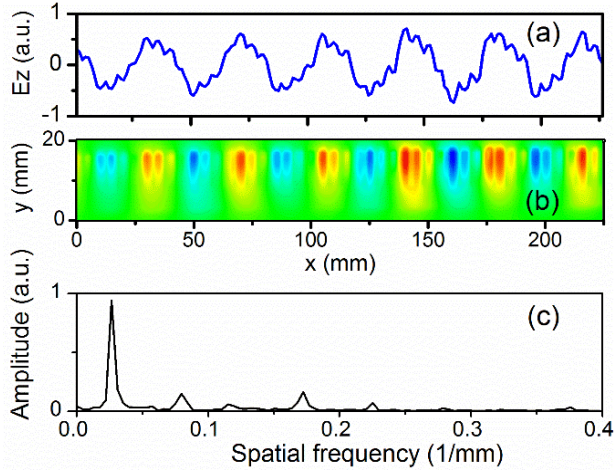

Fig. 3. Field distribution in the real space and its Fourier spectrum. (a) Field profile along *x* axis that have maximum amplitude in *y* direction. (b) Corresponding field distribution in the *x-y* plane. (c) Spatial frequency for the field profile in real space shown in (a). The spatial frequency of the peak amplitude is 1/λ where λ is the wavelength of the corresponding spoof SPP.

With the Fourier spectrum, all the wavenumbers corresponding to spoof SPPs with different frequencies can be obtained. The dispersion curves of the two series of samples are shown in Fig. 4 for comparison. From Fig. 4(a), one can see that in the frequency band up to 20 GHz, the dispersions for aluminum foil sample with *h*=5 mm exhibit only one branch of dispersion curve, while the other samples exhibit two. This is consistent with the theoretical prediction in Eq. (3). For sample with *h*=5 *mm* and *d*=5 *mm*, only the fundamental mode is supported since *h*=*d*. The other samples satisfy the condition that *h*>*d*, thus the 2$^{nd}$ order modes can be supported by these geometric structures. Another phenomenon which can be easily observed in Fig. 4, is that the dispersion curves shift to lower frequencies as the groove depth *h* increases. This is obviously consistent with Eq. (1) and (2). Although Eq. (2) is derived for the surface mode on a periodically corrugated metallic plate, it is still applicable to the ultra-thin version with reasonable deviations.

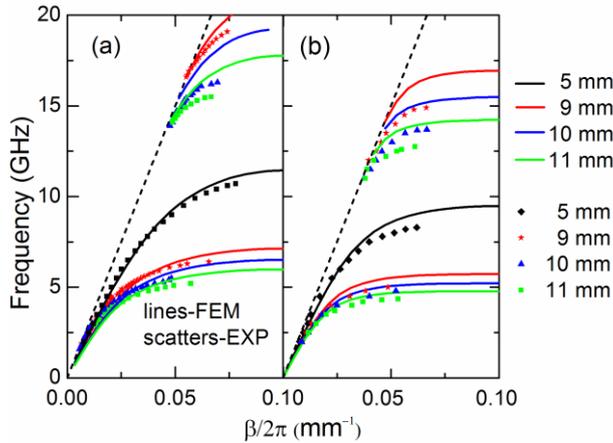

Fig.4. Dispersion curves of spoof SPPs. (a) for aluminum foil samples, and (b) for PCB samples. The groove depths are indicated in the righthand side, and experimental data and FEM simulated results are shown by the scatters and the lines, respectively. The dashed lines represent the light lines. $\beta/2\pi$=0.1 corresponds to the Brillouin zone boundary $\pi/d$.

Comparing with Fig.4 (a), the dispersion curves of PCB strips in Fig. 4(b) display similar behaviors except that all the curves shift to lower frequencies. This red shift is resulted from the higher effective permittivity of the PCB strips due to the substrate effect. In fact, for the PCB strip with $h$=5 mm the 2$^{nd}$ order mode was observed near $f = 20$ GHz, however, this mode is very close to the light line (not shown).

Numerical simulations were also conducted for all the samples and corresponding dispersion curves are drawn in Fig. 4. As a whole, the simulated curves and the experimental ones agree well with each other. For the fundamental modes, simulated curves coincide with the experimental ones quite well; while for the second modes, we note that there're red shift of the experimental results compared to the simulated ones. These red shifts may come from the fact the real experimental situation is more complicated than the simulation conditions. For example, the free standing aluminum foil sample may not be of planar form and the groove surface imperfections can affect the cavity resonance which determines the spoof SPP dispersion. For the fundamental modes, the dispersion is certainly stable in the long wavelength limit even if the structure is perturbed. However, for the higher modes with higher frequencies and wavevectors, the detailed structures including all the imperfections may be resolved by the spoof SPPs. Therefore, the dispersions for the higher modes could be more sensitive to the perturbations.

**Conclusion**

In summary, by using the near filed scanning method, we've measured the electric field distribution of spoof SPPs on two series of ultra-thin comb-shaped metal strips, i.e. PCB and aluminum foil samples. The fundamental modes and 2$^{nd}$ order modes have been observed with distinct field patterns. The wave numbers of spoof SPPs can be obtained by conducting the Fast Fourier transform of field distributions and the dispersion curves have been achieved for various samples. The experimental results are consistent with the theoretical predictions. Numerical simulations agree very well with the experimental results except small red shifts. The near field measurement provide complete information of the fundamental and high order modes and can render further applications possible.

**Acknowledgements**
The author thank Y. Meng, M. D. Yue, P. Y. Cheng, X. T. Bi for sample fabrication. This work is supported by National Natural Science Foundation of China under Grant No. 11304038, Fundamental Research Funds for the Central Universities under Grant No. CQDXWL-2014-Z005 and NSFCQ (cstc2014jcyja50001).

**References**
1. H. Raether, Surface Plasmons, Springer-Verlag, Berlin, 1988.
2. S. A. Maier, Plasmonics: fundamentals and applications, Springer-Verlag, New York, 2007.
3. D. K. Gramotnev, S. I. Bozhevolnyi, Nat. Photonics 4 (2010) 83.
4. M. Fleischmann, P.J. Hendra, A.J. McQuillan, Chem. Phys. Lett. 26 (1974) 163.
5. E. Ozbay, Science 311 (2006) 189.
6. Yu. A. Akimov, H. S. Chu, Nanotechnology 23 (2012) 444004.
7. W. S. Cai, J. S. White, M. L. Brongersma, Nano Lett. 9 (2009) 4403.

8. V. K. Valev. Langmuir 28 (2012) 15454.
9. H. M. Hiepa, T. Endob, K. Kermana, M. Chikaea, D. Kima, S. Yamamuraa,Y. Takamuraa, E. Tamiya, Sci. Tech. Adv. Mater. 8 (2007) 331.
10. J. Zenneck, Ann. Phys. 328 (1907) 846.
11. A. Sommerfeld, Ann. Phys. 333 (1909) 665.
12. J. B. Pendry, L. Martń-Moreno, F. J. Garcia-Vidal, Science 305 (2004) 847
13. R. Ulrich, M. Tacke, Appl. Phys. Lett. 22 (1973) 251.
14. A.P. Hibbins, B. R. Evans, J. R. Sambles, Science 308 (2005) 670.
15. F. J. Garcia-Vidal, L. Martin-Moreno, J. B. Pendry, J. Opt. A, Pure Appl. Opt. 7 (2005) 97.
16. S. A. Maier, S. R. Andrews, L. Martń-Moreno, F. J. Garcń-Vidal, Phys. Rev. Lett. 97 (2006) 176805.
17. C. R. Williams, S. R. Andrews, S. A. Maier, A. I. Fernandez-Dominguez, L. Martin-Moreno, F. J. Garcia-Vidal, Nat. Photonics 2 (2008), 175.
18 X. Shen, T. J. Cui, D. Martin-Cano, F. J. Garcia-Vidal, Proc. Natl. Acad. Sci. U.S.A. 110 (2013) 40.
19. X. Gao, J. H. Shi, X. P Shen, H. F. Ma, W. X. Jiang, L. M. Li, T. J. Cui, Appl. Phys. Lett. 102 (2013)151912.
20. X. Shen and T. J. Cui, Appl. Phys. Lett. 102 (2013) 211909.
21. T. Jiang, L. Shen, X. Zhang, L. Ran, Progress in Electromagnetics Research M 8 (2009) 91.
22. X. Zhang, L. Shen, and L. Ran, J. Appl. Phys. 105 (2009) 013704.
23. X. Y. Liu, Y. J. Feng, B. Zhu, J. M. Zhao, T. Jiang, Opt. Express 21 (2013) 31155.
24. H. Z. Yao S. C. Zhong, Opt. Express 22 (2014) 25149.